\documentclass{tlp}

\usepackage{mathptmx}
\usepackage{amssymb}
\usepackage{graphicx}
\usepackage{wrapfig}
\usepackage{setspace}
\usepackage{multirow}
\usepackage{url}
\usepackage{stmaryrd}
\usepackage{verbatim}
\usepackage{ifthen}
\usepackage{pifont}
\usepackage{tipa}

\usepackage[mathscr]{eucal}

\newcommand{\myfontcodesize}{\fontsize{9}{9}}
\newcommand{\mytt}[1]{{\myfontcodesize \texttt{\textbf{#1}}}}
\newtheorem{proposition}{Proposition}

\newcommand\bcmdtab{\noindent\bgroup\tabcolsep=0pt%
  \begin{tabular}{@{}p{10pc}@{}p{20pc}@{}}}
\newcommand\ecmdtab{\end{tabular}\egroup}

  \title[Applying CLP to SQL Semantic Analysis]%\tnoteref{thanks}]
  {Applying Constraint Logic Programming to\linebreak SQL Semantic Analysis\thanks{Work partially funded by the Spanish Ministry of Economy and Competitiveness, under the grant TIN2017-86217-R (CAVI-ART-2), and by the Madrid Regional Government, under the grant S2018/TCS-4339 (BLOQUES-CM), co-funded by EIE Funds of the European Union. Special thanks are due to the anonymous referees.
  	}
  }

  \author[F. S\'aenz-P\'erez]
         {FERNANDO S\'AENZ-P\'EREZ\\
         Complutense University of Madrid, 28040 Madrid, Spain\\
         \email{fernan@sip.ucm.es}}

\pagerange{\pageref{firstpage}--\pageref{lastpage}}
\newtheorem{definition}{Definition}[section]
\newtheorem{example}{Example}[section]

\begin{document}

\label{firstpage}

\maketitle

\begin{abstract}
This paper proposes the use of Constraint Logic Programming (CLP) to model SQL queries in a data-independent abstract layer by focusing on some semantic properties for signalling possible errors in such queries.
First, we define a translation from SQL to Datalog, and from Datalog to CLP, so that solving this CLP program will give information about inconsistency, tautology, and possible simplifications.
We use different constraint domains which are mapped to SQL types, and propose them to cooperate for improving accuracy.
Our approach leverages a deductive system that includes SQL and Datalog, and we present an implementation in this system which is currently being tested in classroom, showing its advantages and differences with respect to other approaches, as well as some performance data.
This paper is under consideration for acceptance in TPLP.
\end{abstract}


\section{Introduction}
\label{sect:introduction}

Aiding programmers with both syntax and type checking  at compile-time % and other techniques 
obviously improves productivity.
In the realms of SQL, current systems (both proprietary and open-source) typically lack of more advanced techniques such as, in particular, the semantic analysis of statements.
After the syntax checking stage, such an analysis should point out possible incorrect statements (e.g., missing or incorrect tuples in the actual outcome).
%A wrong statement can be either incorrect, because it retrieves more rows than expected, or incomplete, because expected rows are not present in the result.
%Once a statement has been syntactically checked, another higher-level semantic analysis can be performed to detect probably wrong statements \cite{Brass:2006:SES:1183058.1183064}.
%
However, in this paper, we avoid actual execution of statements as done in other approaches (assessment tools, test case generation, data provenance\ldots \cite{c3acdc529d00437980a60fe7f143b86a}), and we target at the compile-time stage instead.

%The motivation of this goal is three-fold: First, learning SQL would be enhanced by presenting not only syntax errors, but also semantic warnings, pointing out to the students suspicious statements.
%Second, acquainted developers would catch semantic errors faster, thus making them more productive.
%And, third, identified errors in this early stage are not transmitted to latter semantic checking of tools as (unit) testing database code and post-mortem analysis, therefore improving the results of these tools.

%This would be seen as a daunting task, but t
There are some indicators of bad statement design which can be used to raise semantic warnings. % to the user of the system.
In particular, we focus on SQL semantic errors as described in \cite{Brass:2006:SES:1183058.1183064} that can be caught independently of the database instance. %(therefore, without resorting to actually solve queries).
There are many possible errors and, among them, the following are included: inconsistent, tautological and simplifiable conditions, uncorrelated relations in joins, unused tuple variables, 
%unnecessary \mytt{DISTINCT} clause, 
constant output columns, duplicate output columns, %comparison with null,
unnecessary general comparison operators, and several others.

%Among all these errors, an interesting case is determining the relevance of conditions.
%Besides proposing how to deal with many of the outlined errors in an actual tool, we focus on the interesting case of determining the relevance of conditions by abstracting the original statements in a logic constraint programming setting with solver cooperation.

Applying such a semantic analysis to SQL is cumbersome because its syntax and semantics do not facilitate expressing program properties \cite{Guagliardo:2017:FSS:3151113.3151116}.
%\cite{DBLP:conf/iccp2/DollingerM11}. 
%Though there are works as \cite{Guagliardo:2017:FSS:3151113.3151116} that formalizes 3-valued SLQ semantics, reasoning at this level is cumbersome.
To ease this task we use Constraint Logic Programming (CLP) \cite{Jaffar:1987:CLP:41625.41635,Apt:2003:PCP:1237975} as a reasoning setting for SQL statements.
This way, we translate an SQL statement into a constraint logic program that in particular models conditions and expressions.
%Each logic variable in the translation represents the possible values its corresponding relation column can take. % for each tuple instance in the context of all the involved relations in the SQL statement.
%Translating includes optimization and simplification techniques as folding/unfolding.
This CLP program is then evaluated to obtain properties of interest for the semantic analysis.
For example, obtaining a failed deduction indicates an inconsistent condition.
%Also, if ground bindings are found, this means that some properties do hold irrespective of the database instance.
%And tautologies can also be easily detected by testing that a complemented condition fails.

CLP systems include different solvers for specific constraint domains such as Booleans, finite domains, reals, and rationals.
Each one is an instance of the generic schema CLP($\mathscr{X}$),  %\cite{Jaffar:1987:CLP:41625.41635,Apt:2003:PCP:1237975}, 
where $\mathscr{X}$ is a constraint domain which can be mapped to an SQL type.
On the one hand, since a \mytt{WHERE} condition generally includes columns of different types, then different domains (and, therefore, constraint solvers) are expected to be involved in a single condition. % (that is, mapping types to domains).
On the other hand, the deduction power of each solver is limited by its constraint propagators and the kind of constraints it can deal with. 
For example, while a finite domain solver can handle non-linear constraints, a real solver cannot.
Thus, we apply \textit{solver cooperation} \cite{HofstedtCP:2000} to enable solver cooperation for compatible domains and interchange deductions to improve accuracy.
%So, we adapt this technique to our system to get improved deductions.

We have implemented our proposal in a deductive database system that includes %(among others) 
SQL as a query language.
This system (Datalog Educational System -- DES \cite{saenzDESentcs11}) is an interactive tool mainly targeted at teaching, and it is appealing for SQL learning with the aid of both syntax and semantic checking (as presented here).
It has experienced more than 76K downloads and has been used in more than 50 universities around the world (cf. {\myfontcodesize\url{des.sourceforge.net/html/facts.html}}).
Solving a query %of the different query languages it supports 
is via an optimized translation into a Datalog program, which is then solved by its deductive engine.
Thus, we take advantage of this Datalog translation for the generation of a CLP program.
To the best of our knowledge, this is the first work dealing with SQL semantic errors using CLP.

We are currently using the system for our \textit{Databases} modules via a web interface ({\myfontcodesize\url{desweb.fdi.ucm.es}}), retrieving data to evaluate the usefulness of the semantic warnings.
More than 200 student accounts have been created, and more than 3,000 logins have been registered, including 600 guest account logins.
%
%Our main contribution is, thus, to provide a comprehensive system including the outlined approach.
%%The system not only includes semantic checking, but also detailed syntax checking (more than several commercial DBMS's as also shown in Section \ref{sect:the-system}).
%The closest related work \cite{Brass:2006:SES:1183058.1183064} employs satisfiability tests and model construction to generate warnings that are not always accurate and deduce less information, as it will be shown in Section \ref{sect:related-work}.
%A more detailed related work study is conducted in Section \ref{sect:related-work}.
Next, the proposal is motivated by examples.

%\subsection{Motivating Examples}
%\label{sect:motivation}

\paragraph{Motivating Examples}Following \cite{Brass:2006:SES:1183058.1183064}, a simple semantic error occurs in the following query:%\footnote{Keywords in SQL code shown in this paper are capitalized, though SQL is case insensitive in most systems.}

{\myfontcodesize
\begin{verbatim}
SELECT * FROM employees WHERE dept='IT' AND dept='HR';
\end{verbatim}
}\vspace{-2mm}

\noindent Here, the condition is trivially false due to (probably) using the wrong logical operator.
Despite this, it is accepted and solved with no warning in current DBMSs.

Conditions also appear in database constraints, %\footnote{Not to be confused with constraints in the CLP setting.} 
and may be identified as either inconsistent or tautological.
Consider the following definitions, in which the constraint on the salary has the minimum and maximum values interchanged (no definite tuple could ever be inserted): % in such a table with a non-null salary value):

{\myfontcodesize
\begin{verbatim}
CREATE TABLE departments(dept VARCHAR(10) PRIMARY KEY, dname VARCHAR(20));
CREATE TABLE employees(ename VARCHAR(20), dept VARCHAR(10) REFERENCES departments, 
                       salary INT CHECK salary BETWEEN 5000 AND 2000);
\end{verbatim}
}\vspace{-3mm}

Tautological conditions can occur as in the following statement (where the intention would probably be to use \mytt{AND} instead of \mytt{OR}):

{\myfontcodesize
\begin{verbatim}
CREATE TABLE employees(ename VARCHAR(20), dept VARCHAR(10) REFERENCES departments, 
                       salary INT CHECK salary > 2000 OR salary < 5000);
\end{verbatim}
}\vspace{-3mm}

We can consider a more involved example including both database constraints in a table and an SQL query.
On it, a table is defined for containing gas products and describing their composition as percentages, which must make a total of one hundred percent.
%This amounts to a reasonable set of database constraints.
The \mytt{SELECT} query below would be inconsistent because it is asking for a gas product with components summing more than 100\%.

{\myfontcodesize
\begin{verbatim}
CREATE TABLE gas_products(name      VARCHAR(20) PRIMARY KEY, 
                          butane    FLOAT CHECK butane    BETWEEN 0 AND 100, 
                          propane   FLOAT CHECK propane   BETWEEN 0 AND 100, 
                          olefins   FLOAT CHECK olefins   BETWEEN 0 AND 100, 
                          diolefins FLOAT CHECK diolefins BETWEEN 0 AND 100, 
                          CHECK     butane+propane+olefins+diolefins =  100);

SELECT name FROM gas_products WHERE butane>60 AND propane>50;
\end{verbatim}
}

Finally, another possibility is a condition that can be simplified, which may be a symptom of a wrong condition.
For example:

{\myfontcodesize
\begin{verbatim}
SELECT butane, propane FROM gas_products 
WHERE butane-propane=10 AND butane+propane=80;
\end{verbatim}
}

This is equivalent to the simple condition \mytt{butane=45 AND propane=35} because the condition represents a system of linear equations with a single solution. 
Then, both output columns are constants, and therefore symptoms of a wrong query.
%In this case, it may be the case that the pattern of the condition is correct (i.e, looking for a given difference between components and a given amount of two of them) but perhaps an incorrect column name has been used. % because of a copy and paste error.
%
Other errors that students typically make and are also covered by this tool %(such as detecting uncorrelated relations, unneeded relations\ldots
are presented in Subsection \ref{sect:errors}.
% but are out of the scope of this paper.
%Another common error is to forget the correlation in joins, as in:
%
%{\myfontcodesize
%\begin{verbatim}
%SELECT ename, dname FROM employees, departments;
%\end{verbatim}
%}
%
%This error is easily caught at first sight, but in a complex query including many relations and conditions, it might not be so simple, so that an automated tool is an advantage.
%
%Similar to this example, in the following one there is no need to include the relation \mytt{departments} (recall the foreign key in the column \mytt{employees.dept}):
%
%{\myfontcodesize
%\begin{verbatim}
%SELECT ename, dept FROM employees NATURAL INNER JOIN departments;
%\end{verbatim}
%}

%Such errors are typically faced not only by students learning SQL (\cite{Brass:2006:SES:1183058.1183064} and also in our personal experience along teaching databases during many years), but also by programmers with enough skills.

%:: UPDATE examples.pdf in www.fdi.ucm.es

Next sections detail our approach to identify such wrong uses of SQL conditions, and consists in translating an SQL statement into a CLP program, which is evaluated for identifying inconsistency, tautology, simplifiable conditions, and constant output columns.
% and several other errors as summarized in Subsection \ref{sect:errors}. 
Since we use DES, which translates SQL to Datalog, we start from this translation (Section \ref{sect:sql-to-datalog}) for generating the CLP program (Section \ref{sect:datalog-to-clp}).
Section \ref{sect:the-system} presents the working system with the techniques used to identify a collection of semantic errors, together with performance data.
Section \ref{sect:related-work} relates this work to other approaches and, finally, we present in Section \ref{sect:conclusions} our conclusions and points for future work.

\section{From SQL to Datalog}
\label{sect:sql-to-datalog}

In a first stage, we take advantage of the translation from an SQL query to a semantically equivalent Datalog program.
It builds upon the basic presentation in \cite{Ullman88}, and extended in \cite{Sae17a} (where formal results for semantic equivalence are given).
%So, we assume the function \mytt{{\em SQL\_to\_DL}}($r$, \mytt{$Q$}) (and, for the sake of space, we refer the reader to its partial description in \cite{Sae17a}), which takes a relation name $r$ and an SQL query $Q$ defining a relation as input, and returns a multiset of Datalog rules providing the same meaning as the SQL relation for a corresponding predicate with the same name as the relation. 
First, we specify the syntax of the language fragments we consider for both SQL and Datalog, then we describe the translation, and finally some examples are presented.

\subsection{SQL Syntax}
In this section, we consider a fragment of standard SQL \cite{sql:2016}.
% as found in many textbooks as \cite{Silberschatz6th}.
Despite our approach supports a wider coverage of SQL than the considered here, we stick to the grammar in Figure \ref{fig:sql-syntax} for the sake of simplicity.
There, {\tt true-typed} words stand for terminal symbols, `c' for constants, `r' for relations (either tables or views), and `a' for relation attributes. 
In this grammar (and the one in the next subsection), we use the symbol \mytt{::=} for defining parts of the language, square brackets (\mytt{[ ]}) to delimit optional parts, and vertical bars (\mytt{|}) to separate alternative parts.
%Set operators remove duplicates and the keyword \mytt{ALL} is omitted for simplicity.
We assume a type inference system for syntactically valid queries.
Also, we assume that syntax comprehensions such as \mytt{E BETWEEN E1 AND E2} are re-written in their equivalent basic forms supported by the grammar. %; in this case it would have been rewritten as \mytt{E>=E1 AND E<=E2}.
%\footnote{For only the core language, the current SQL standard spans hundreds of pages in two volumes.} including no correlations.
Each relation alias \mytt{Relation AS Alias} in a \mytt{FROM} clause is re-written as a reference to the alias, by adding a new relation \mytt{Alias} $\leftarrow$ \mytt{Relation} to the database, where the symbol $\leftarrow$ stands for relation definition.

\begin{figure}[tb]
%\noindent\rule{\linewidth}{0.4pt}
		\begin{tabbing}%\vspace*{-4mm}
%\noindent\rule{\linewidth}{0.4pt}\\
			$query$  \= ::= \= \mytt{SELECT [ALL|DISTINCT] [TOP} c \mytt{]} $exp$,...,$exp$ \mytt{FROM} $rel$,...,$rel$ [ \mytt{WHERE} $cond$ ] $\mid$ \\
			\>\> $query$ $s\_op$ \mytt{[ALL|DISTINCT]} $query$ \\
%			&     & \mytt{WITH} R \mytt{AS} $query$ $query$ \\
			$exp$  \> ::= \> c $\mid$ r.a $\mid$ $exp$ $m\_op$ $exp$ $\mid$ $-exp$ $\mid$ $query$ \\
			$rel$  \> ::= \> r $\mid$ $query$ \\ 
			$cond$  \> ::= \> $exp$ $c\_op$ $exp$ $\mid$ \mytt{NOT} $cond$ $\mid$ $cond$ $l\_op$ $cond$ $\mid$  \mytt{TRUE} $\mid$ \mytt{FALSE} $\mid$ 

			 $exp$ \mytt{IN} $query$ $\mid$ \mytt{EXISTS} $query$\\
			$c\_op$ \> ::= \> \mytt{>} $\mid$ \mytt{<} $\mid$ \mytt{=} $\mid$ \mytt{<>} $\mid$ \mytt{>=} $\mid$ \mytt{<=} \\
			$m\_op$ \> ::= \> \mytt{+} $\mid$ \mytt{-} $\mid$ \mytt{*} $\mid$ \mytt{/} \\
			$s\_op$ \> ::= \> \mytt{UNION} $\mid$ \mytt{EXCEPT} $\mid$ \mytt{INTERSECT}\\
			$l\_op$ \> ::= \> \mytt{AND} $\mid$ \mytt{OR}
		\end{tabbing}
	\vspace{-5mm}
\noindent\rule{\linewidth}{0.4pt}
	\caption{A grammar for a subset of standard SQL}
	\label{fig:sql-syntax}
\end{figure}

A query can appear directly as a row-returning SQL statement, as well as in other statements of the DML (Data Manipulation Language) such as \mytt{INSERT} and \mytt{DELETE} statements.
For example, in: \mytt{INSERT INTO} r $query$ (where the results of $query$ are inserted into the relation `r').
DDL (Data Definition Language) statements such as \mytt{CREATE TABLE} can include predicates (following the syntax of $cond$ in Figure \ref{fig:sql-syntax}) in \mytt{CHECK} constraints. 
Note that both conditions and expressions can include queries as it can be seen in the definition of $cond$ and $exp$, respectively.
The DDL statement \mytt{CREATE VIEW AS} $query$ also includes a query (following the syntax of $query$ in the same figure).
Thus, queries and conditions occurring in any part of an SQL statement are targets for the proposed semantic analysis.

\vspace{-3mm}
\subsection{Datalog Syntax}
With respect to Datalog, we consider an extended Datalog language with duplicates and metapredicates as shown in Figure \ref{fig:datalog-syntax},  where $rule$ stands for rules, $goal$ for goals, $exp$ for expressions, `atom' for an atom  (possibly containing variables and constants), the comma (`,') for a conjunction, and the semi-colon (`;') for a disjunction.
$c\_op$ and $m\_op$ are the same as in Figure \ref{fig:sql-syntax} excepting \mytt{<=}, which is written as \mytt{=<}. % to avoid operator clash with the implication to the left.
The syntax of the logic includes a universe of constant symbols, a set of variables, a set of user-defined predicates, and a set of built-in metapredicate symbols (where the prefix operator \mytt{not} stands for negation, the predicate \mytt{distinct}/1 for duplicate elimination, and \mytt{top}/2 for the first $n$ solutions of a goal).
%($\mathscr{P}$).
Following Prolog syntax, variables are written starting with either an upper-case letter or an underscore, and the rest of symbols either starting with lower-case or delimited by single quotes.
%Removing function symbols from the logic is a condition for finiteness of answers, a natural requirement of relational database users.
%As in Horn-logic, a rule has the form $A \mytt{:-} \phi$, where $A$ is an atom and $\phi$ is a conjunction of goals.
%Since we consider a hypothetical system,
%a goal can also take the form $R_1 \land \ldots \land R_n \Rightarrow G$, a construction known as an {\em embedded implication}.
%As an embedded implication behaves different from a regular implication \cite{bonner90adding}, it receives a different syntax symbol: $\Rightarrow$.
%The following definition captures the syntax of the language, where $vars(T)$ is the set of variables occurring in $T$:
%
%
The first form of $rule$ in the figure is also known as a fact. % and it represents a true knowledge of the atom, whereas a rule of the second form represents that the atom is considered to be true if the goals to the right of `\mytt{:-}' (implication to the left) can be deduced to be true.
\setlength{\tabcolsep}{1pt}
\begin{figure}[b]
%	  \begin{center}
\noindent\rule{\linewidth}{0.4pt}
%	\begin{minipage}[b]{\linewidth}
%	  \begin{center}
%       \fbox{
		\begin{tabular}{lcl}
			
			$rule$ & ::= & atom $\mid$ 
			 atom \mytt{:-} $goal$ \mytt{,} \ldots \mytt{,} $goal$ \\
			$goal$ & ::= & atom $\mid$ \mytt{not} atom $\mid$ \mytt{distinct}(atom) $\mid$ \mytt{top}(c, atom) $\mid$ 
            $goal$ \mytt{;} \ldots \mytt{;} $goal$ $\mid$ 
            $exp$ $c\_op$ $exp$ \\
			$exp$ & ::= & \mytt{X} $\mid$ $exp$ $m\_op$ $exp$ $\mid$ $-exp$ \\
			
		\end{tabular}
%		}
%	  \end{center}
%	\end{minipage}
\noindent\rule{\linewidth}{0.4pt}
%\end{center}
  \caption{A grammar for an extended Datalog language}
  \label{fig:datalog-syntax}
\end{figure}
A Datalog database (also referred to as a program) contains facts and rules instead of relation definitions as in SQL (tables and views, respectively).
We consider also a type system for Datalog for restricting valid rules with respect to type specifications.

\subsection{Translation}
\label{sect:sql-to-dl-translation}

This section describes some examples of the translation of the considered SQL and Datalog languages, extending the description in \cite{Sae17a} with the clause \mytt{DISTINCT} and the operators \mytt{IN} and \mytt{EXISTS}.
%In order to process SQL statements, the deductive system DES resorts to compile such statements to Datalog rules and queries \cite{DBLP:conf/aplas/Saenz-PerezCG11}.
Here, we refer to the function \mytt{{\em SQL\_to\_DL}} as defined there (which we do not reproduce it here).
It takes a relation name and an SQL query defining a relation as input, and returns a multiset of Datalog rules providing the same meaning as the SQL relation for the corresponding predicate with the same name as the relation.
For a query in the top-level, we assign a relation name (\mytt{answer}) to build the outcome. %, similarly to other database systems.
%The following definition for this function includes only a couple of the basic cases, where others can be easily developed from \cite{Ullman88}.
From here on, set-related operators and symbols refer to multisets because SQL relations can contain duplicates. %(Note that Hypothetical Datalog as defined in \cite{sae13c-ictai13} deals with duplicates as well.)

%\begin{definition}
%	\noindent \\

An SQL query is preprocessed before passing it to the translation function:

\begin{itemize}
	\item 		
	If the keyword \mytt{DISTINCT} is specified in a query $Q$ defining a relation $r$, this query is re-written as follows, where a fresh relation $r'$ is introduced, and the notation $Q[X/Y]$ means a syntactic replacement of $X$ by $Y$ in $Q$:
	
	\medskip
	
	\noindent $r \leftarrow$ \mytt{SELECT DISTINCT * FROM} $r'$ \qquad
	 $r' \leftarrow$ $Q[$\mytt{DISTINCT}$/$\mytt{ALL}$]$ 
	
	\medskip
	
	\noindent 
	Note that a \mytt{SELECT} statement without a \mytt{WHERE} clause means an implicit \mytt{true} condition.
	
	\item 		
	If a set operator includes (either implicitly or explicitly) the keyword \mytt{DISTINCT} in a query $Q \equiv Q_1 ~ s\_op ~ Q_2$ defining a relation $r$, then it is re-written as:
	
	\medskip
	
	\noindent $r \leftarrow$ \mytt{SELECT DISTINCT * FROM} $r'$  \qquad
	$r' \leftarrow$ $Q_1 ~ s\_op$ \mytt{ALL} $Q_2$ 
	
\end{itemize}

In addition, we define a function to deal with a set of SQL relation definitions which can appear as the result of SQL preprocessing:

\begin{definition} \label{def:SQLstoDL} The function \mytt{{\em SQLs\_to\_DL}} takes a set of SQL relation definitions as input and returns the equivalent Datalog program:
\mytt{{\em SQLs\_to\_DL}}($\{ r_1 \leftarrow SQL_1, \ldots , r_n \leftarrow SQL_n\}$) =  $\bigcup\limits_{i=1}^{n}$\mytt{{\em SQL\_to\_DL}}($r_i, SQL_i)$\hfill $\square$

\end{definition}

\begin{example}
\label{ex:ex1}
	Given the following table schemas:
    
%    \medskip
    
\noindent
\mytt{CREATE TABLE dept(id CHAR(10) PRIMARY KEY, name CHAR(20), location CHAR(20))};\\
\mytt{CREATE TABLE emp(name CHAR(20) PRIMARY KEY,}\\
\hspace*{2.82cm}\mytt{dept CHAR(10) REFERENCES dept(id), salary INT);}

%    \medskip
    
    And a query in the top level that lists the employee names and their department names:
    
%    \medskip

	\mytt{answer} $\leftarrow Q$\\ \indent 
    $Q \equiv$ \mytt{SELECT emp.name, dept.name FROM emp, dept WHERE emp.dept=dept.id}
    
%    \medskip

    Since it is a single query definition, we apply the function \mytt{{\em SQL\_to\_DL}} to obtain:

%    \medskip

    \begin{tabbing}
    \mytt{{\em SQL\_to\_DL}}(\mytt{answer}, $Q$) = \= 
    $\{$(\mytt{answer(X$_1$,X$_5$)} \mytt{:-} \mytt{r$_1$(X$_1$,X$_2$,X$_3$,X$_4$,X$_5$,X$_6$)}\mytt{,} \mytt{true}\mytt{,} \mytt{true}\mytt{,} \mytt{X$_2$=X$_4$})$\} \bigcup$ \\
    \> $\{$(\mytt{r$_1$(X$_1$,X$_2$,X$_3$,X$_4$,X$_5$,X$_6$)} \mytt{:-} \mytt{emp(X$_1$,X$_2$,X$_3$}), \mytt{dept(X$_4$,X$_5$,X$_6$)})$\}$
    \end{tabbing}\vspace{-0.55cm}\hfill $\square$

%	\indent where:\\
%	\mytt{{\em SQLREL\_to\_DL}}((\mytt{emp, dept})) = (\mytt{r$_1$(X$_1$,X$_2$,X$_3$,X$_4$,X$_5$,X$_6$)}, $\{$(\mytt{r$_1$(X$_1$,X$_2$,X$_3$,X$_4$,X$_5$,X$_6$)} \mytt{:-} \mytt{emp(X$_1$,X$_2$,X$_3$}), \mytt{dept(X$_4$,X$_5$,X$_6$)})$\}$), \\
%    \mytt{{\em SQLCOND\_to\_DL}}(\mytt{emp.dept=dept.id}) = ((\mytt{X$_2$=X$_4$}), $\emptyset$), \\
%	\mytt{{\em SQLEXP\_to\_DL}}(\mytt{emp.name}) = (\mytt{X$_1$}, \mytt{true}, $\emptyset$), and \\
%	\mytt{{\em SQLEXP\_to\_DL}}(\mytt{dept.name}) = (\mytt{X$_5$}, \mytt{true}, $\emptyset$). \\
    
\end{example}

\begin{example}
	Given the same table schemas as in the previous example, a query in the top-level that lists the department names with assigned employees is
	\mytt{answer} $\leftarrow Q$, where:
	 
	\indent 
	$Q \equiv$ \mytt{SELECT dept.name FROM dept WHERE dept.id  IN} \\
	\indent \indent \indent \mytt{ (SELECT DISTINCT dept.name FROM emp, dept WHERE emp.dept=dept.id)}
	
	\medskip
	
	This is re-written as \mytt{answer} $\leftarrow Q_1$ and \mytt{r$_2$} $\leftarrow Q_2$, where:
	
	\indent
	$Q_1 \equiv$ \mytt{SELECT dept.name FROM dept WHERE dept.id IN SELECT DISTINCT * FROM r$_2$} \\ \indent
	$Q_2 \equiv$ \mytt{SELECT dept.name FROM emp, dept WHERE emp.dept=dept.id}

	\medskip

%	$Q \equiv$ \mytt{SELECT dept.name FROM dept WHERE dept.id NOT IN} $Q'$\\ \indent
%	$Q' \equiv$ \mytt{SELECT DISTINCT * FROM p}\\ \indent
%%	\mytt{p'} $\leftarrow$ $Q'$ \\ \indent
%	\mytt{p} $\leftarrow$ \mytt{SELECT dept.name FROM emp, dept WHERE emp.dept=dept.id}
%	
%	\medskip
	
	Since there are two relation definitions, we use Definition \ref{def:SQLstoDL}:

	\medskip

	\mytt{{\em SQLs\_to\_DL}}($\{$\mytt{answer} $\leftarrow Q_1$, \mytt{r$_2$} $\leftarrow Q_2 \}$) = \mytt{{\em SQL\_to\_DL}}(\mytt{answer}, $Q_1$) $\bigcup$
	\mytt{{\em SQL\_to\_DL}}(\mytt{r$_2$}, $Q_2$) 
	
	\medskip

	$Q_2$ is almost identical to $Q$ in Example \ref{ex:ex1}; the  differences are an absent argument in the projection, and different names for relations and variables.
%	So, we omit its translation steps.
	The translation of \mytt{answer} $\leftarrow Q_1$
    %and denoting with $Q'$ the subquery after \mytt{NOT IN} in $Q_1$, 
    is:
	
%	\medskip

\begin{tabbing}
	\mytt{{\em SQL\_to\_DL}}(\mytt{answer}, $Q_1$) \= = \= 
	$\{$(\mytt{answer(X$_2$)} \mytt{:-}  \mytt{dept(X$_1$,X$_2$,X$_3$)}\mytt{,} \mytt{true, X$_4$=X$_1$, r$_1$(X$_4$)})$\} \bigcup \emptyset \bigcup$ \\
	\> \> $\{$(\mytt{r$_1$(X$_4$) :- distinct(r$_2$(X$_4$))})$\}$, and then:
\end{tabbing}
	
%	$\bigcup $
%	(\mytt{r(X$_8$)} \mytt{:-} \mytt{emp(X$_4$,X$_5$,X$_6$})\mytt{,} \mytt{dept(X$_7$,X$_8$,X$_9$)}\mytt{,} \mytt{X$_5$=X$_7$}\mytt{,} \mytt{true}\mytt{,} \mytt{true})
	
	%\medskip
	
%	\indent where:
%	
%	\medskip
%	
%    \noindent
%	\mytt{{\em SQLREL\_to\_DL}}(\mytt{dept}) = (\mytt{dept(X$_1$,X$_2$,X$_3$)}, $\emptyset$), \\
%	\mytt{{\em SQLCOND\_to\_DL}}(\mytt{dept.id NOT IN} $Q'$) = ((\mytt{true, X$_4$=X$_1$, not r$_1$(X$_4$)}), $\{$(\mytt{r$_1$(X$_4$) :- distinct(r$_2$(X$_4$))})$\}$), \\
%	\mytt{{\em SQLEXP\_to\_DL}}(\mytt{dept.id}) = (\mytt{X$_1$}, \mytt{true}, $\emptyset$), \\
%	\mytt{{\em SQLREL\_to\_DL}}($Q'$) = $($ \mytt{r$_1$(X$_4$)}, $\{($\mytt{r$_1$(X$_4$) :- distinct(r$_2$(X$_4$))})$\}  )$, \\  
%	\mytt{{\em SQL\_to\_DL}}(\mytt{r$_1$}, $Q'$) = 
%	$\{$(\mytt{r$_1$(X$_4$)} \mytt{:-}  \mytt{distinct(r$_2$(X$_4$))})$\}$
%
%	\medskip

%	Then:
%	
%	\medskip

\begin{tabbing}
	\mytt{{\em SQLs\_to\_DL}}\=($\{$\mytt{answer} $\leftarrow Q_1$, \mytt{r$_2$} $\leftarrow Q_2 \}$) = \\%(($\{$\mytt{answer} $\leftarrow Q_1$, \mytt{r$_1$} $\leftarrow Q_2 \}$) = \\
    \> $\{$\=(\mytt{answer(X$_2$)} \mytt{:-}  \mytt{dept(X$_1$,X$_2$,X$_3$)}\mytt{,} \mytt{true, X$_4$=X$_1$, r$_1$(X$_4$)}), \\
	\> \> (\mytt{r$_1$(X$_4$) :- distinct(r$_2$(X$_4$))}), \\	
	\> \> (\mytt{r$_2$(X$_8$)} \mytt{:-} \mytt{r$_3$(X$_5$,X$_6$,X$_7$,X$_8$,X$_{9}$,X$_{10}$)}\mytt{,} \mytt{true}\mytt{,} \mytt{true}\mytt{,} \mytt{X$_6$=X$_8$}), \\
    \> \> (\mytt{r$_3$(X$_5$,X$_6$,X$_7$,X$_8$,X$_{9}$,X$_{10}$)} \mytt{:-} \mytt{emp(X$_5$,X$_6$,X$_7$)}, \mytt{dept(X$_8$,X$_{9}$,X$_{10}$)})$\}$
\end{tabbing}\vspace{-0.8cm}\hfill $\square$
\end{example}

In these simple examples, the generated Datalog program can be simplified by removing \mytt{true} goals and explicit variable bindings (e.g., \mytt{X$_4$=X$_1$}), and by applying substitutions (a goal \mytt{X=X} is a trivially \mytt{true} goal and thus it can also be removed).
In addition, folding/unfolding techniques %\cite{Sterling:1994:APA:175753} 
\cite{Burstall:1977:TSD:321992.321996,ts84}
are applied to further simplify the Datalog program generated.
Here, unfolding can be applied to user predicate calls for which the predicate consists of only one clause, thus removing the predicate itself.
Following Example \ref{ex:ex1}, the translated program is simplified into the following single rule (with a substitution [\mytt{X$_4$}/\mytt{X$_2$}]):\\
$\{$(\mytt{answer(X$_1$,X$_5$)} \mytt{:-} \mytt{emp(X$_1$,X$_2$,X$_3$}), \mytt{dept(X$_2$,X$_5$,X$_6$)})$\}$.

%Sometimes, we stumble into contrived SQL formulations.
%An equivalent, but cumbersome, query for Example \ref{ex:ex1} is: 
%
%\mytt{SELECT e.name, d.name} \\ \indent 
%\mytt{FROM (SELECT * FROM emp) AS e, (SELECT * FROM dept) AS d WHERE e.dept=d.id}

\section{From Datalog to CLP}
\label{sect:datalog-to-clp}

Once the Datalog program corresponding to an SQL query is obtained, we can reason in the logical level about program properties of interest.
Selecting a CLP language for expressing such properties seems to be a natural choice for dealing with unbound variables in conditions.
Note that SQL conditions operate on data coming from their providers (relation instances, either table contents or the result of solving a reference to a view).
However, our compile-time approach avoids inspecting such instances, thus leading to non-ground conditions in general.
If all the conditions are expressed as a constraint problem, solvers can be used to infer some deductions.

For example, the SQL condition \mytt{r.a > s.b AND s.b > r.a} is translated into its Datalog equivalent \mytt{X > Y, Y > X}.
Since \mytt{X} and \mytt{Y} are logic variables, the condition cannot be tested by a Datalog deductive engine without resorting to retrieve data from the database instance (this would make the condition ground for concrete data in safe rules \cite{Ullman88}).
However, it can be posted to a solver which could deduce an inconsistent state with the help of its constraint propagators.
Similarly, a tautological condition can be identified by determining whether its complement is false, as in the example in the introduction: \mytt{salary INT CHECK salary > 2000 OR salary < 5000}.
The condition would be translated into \mytt{X > 2000; X < 5000} and a constraint solver can deduce that its complement \mytt{X =< 2000, X >= 5000} is false. 
Such a solver can also simplify its constraint store.
For example, given the following condition posed in the introduction: \mytt{butane-propane=10 AND butane+propane=80}, its translation \mytt{X-Y=10, X+Y=80} forms a conjunction of linear equations for which a numeric solver can find the single solution \mytt{X=45} and \mytt{Y=35}.

Therefore, our approach consists in translating the Datalog program into a CLP program in which Datalog conditions are replaced by CLP constraints.
Moreover, since this CLP reasoning operates at compile-time, deductions are independent of database instances.
Nonetheless, as each variable occurring in the CLP program %(which is universally quantified by default) 
has attached the domain of all possible values for its type, each base relation (table) in the Datalog program is translated with the CLP constraints corresponding to the \mytt{CHECK} constraints.
%so that the domain of the variable is not restricted by its concrete data provider instance and becomes universally quantified.
This way, solving the CLP program represents an abstract solving of the original Datalog program. 
%Summarizing the proposed stages

Next subsection identifies both SQL data types and their corresponding constraint domains.
Since there can be different solvers for compatible domains (with different deduction capabilities), in Subsection \ref{sect:cooperation} we show how they interact to improve deductions via cooperation.

\subsection{Domains}

Constraint solvers operate on specific constraint domains, which we map to compatible SQL data types. 
SQL standard data types 
%\cite{sql:2016} 
include in particular exact numeric (\mytt{INTEGER}, \mytt{NUMERIC}, \mytt{DECIMAL}, \mytt{SMALLINT}), approximate numeric (\mytt{FLOAT}, \mytt{REAL}, \mytt{DOUBLE PRECISION}), Boolean (\mytt{BOOL}), character string (\mytt{CHAR}, \mytt{VARCHAR}), and datetime types (\mytt{DATE}, \mytt{TIME}, \mytt{TIMESTAMP}).

Constraint domains are instances of the generic schema CLP($\mathscr{X}$) \cite{Jaffar:1987:CLP:41625.41635,Apt:2003:PCP:1237975}, where $\mathscr{X}$ is a constraint domain.
Typically, the following domains can be found in existing implementations: $\mathscr{FD}$ (finite domain of integers, with both linear and non-linear constraints), $\mathscr{Q}$ (rational numbers with linear constraints), approximate numeric ($\mathscr{R}$), and Boolean ($\mathscr{B}$).

Further constraint solvers can be developed with the aid of Constraint Handling Rules (CHR) %\cite{Frhwirth:2009:CHR:1618539}
which eases the task of implementing specific-application constraints.
In particular, string solvers (which have received recently a large amount of research \cite{6735366,CI15}) could be developed over character strings, and so do solvers over datetime types.
 
\subsection{Solver Cooperation}
\label{sect:cooperation}

Solver cooperation \cite{CES18a,DBLP:journals/corr/abs-0904-2136,HofstedtCP:2000,DBLP:conf/advis/MonfroyC04} is a technique enabling solver interaction with the aim to early prune the search space during solving.
This is possible because different solvers prune the search space in different ways.
For example, on the one hand, given the constraints \mytt{X+Y=2} and \mytt{X-Y=0}, the propagators of the CLP($\mathscr{FD}$) solver are typically not able to solve the linear system, though the CLP($\mathscr{Q}$) does.
% \mytt{X+Y=2} and \mytt{X-Y=0}: SICStus (no), SWI-Prolog(yes), GNU-Prolog (no), Ciao(?), XSB (?), Eclipse (no)
On the other hand, considering the constraints \mytt{X*X=4}, \mytt{X>0} and \mytt{X<4}, the CLP($\mathscr{FD}$) solver is able to solve the non-linear problem, whereas CLP($\mathscr{Q}$) does not.
Thus, provided that CLP systems enjoy different solvers, we take advantage of them and make them to cooperate.

Logic programming (LP) systems include the Herbrand domain $\mathscr{H}$ that supports computations with symbolic equality and disequality constraints over values of any type. 
In the CLP setting, the domain $\mathscr{H}$ can directly cooperate with other solvers wherever each variable is attached to a single CLP solver (in order to prevent domain clash).

Each condition occurring in the Datalog program is translated into a constraint term which includes the target domain type.
With this indication, the constraint is sent to the corresponding solver(s) at CLP evaluation time.
If a single solver $\mathscr{X}$ is available for a given domain type, then the constraint operates on the same logic variables occurring in the translated CLP program (i.e., a direct cooperation of $\mathscr{H}$ with $\mathscr{X}$).
Otherwise, when more than a solver is available (on several domains $\mathscr{X}_i$), a copy of each  variable in the constraint is created for each solver, and bridge constraints are imposed to make possible the bidirectional communication between them.

\begin{figure}[b]
  %\begin{center}
%    \begin{minipage}{0.95\linewidth}
  \fbox{
           \includegraphics[width=0.97\linewidth]{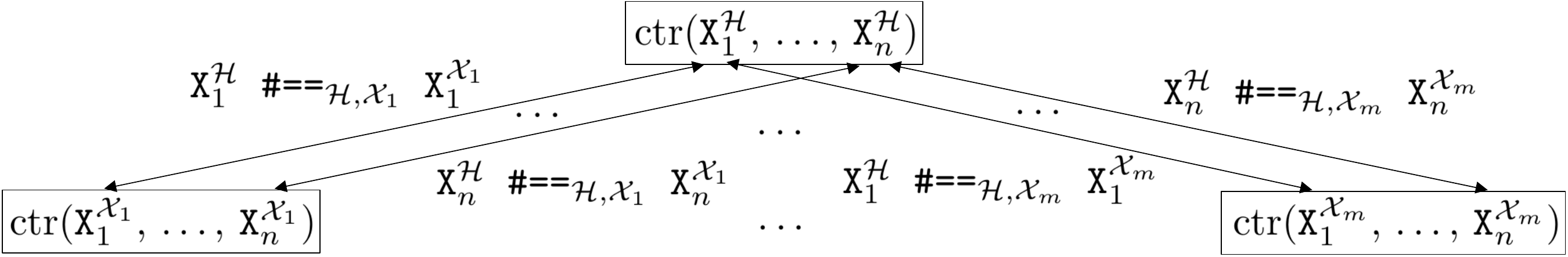}
    }
%    \end{minipage}
  %\end{center}       
\caption{Solver Cooperation} 
\label{fig:solver-cooperation}
\end{figure}

Figure \ref{fig:solver-cooperation} illustrates our approach to solver cooperation (which differs from \cite{DBLP:journals/corr/abs-0904-2136} since we do not take projections into account).
A bridge constraint is denoted as \mytt{X}$^{\mathscr{X}_{1}}$ \mytt{\#==}$_{\mathscr{X}_{1},\mathscr{X}_{2}}$ \mytt{X}$^{\mathscr{X}_{2}}$, which relates two variables \mytt{X}$^{\mathscr{X}_{1}}$ and \mytt{X}$^{\mathscr{X}_{2}}$ in the domains $\mathscr{X}_{1}$ and $\mathscr{X}_{2}$, respectively.
In the figure, the constraint ctr(\mytt{X$_1^{\mathscr{H}}$}, \ldots, \mytt{X$_n^{\mathscr{H}}$}) operating on the domain $\mathscr{H}$ has $n$ variables \mytt{X}$_i^{\mathscr{H}}$ ($1 \leq i \leq n$), and there are $m$ compatible solvers for which $m$ equivalent $\mathscr{X}_{j}$ constraints ctr(\mytt{X$_1^{\mathscr{X}_{j}}$}, \ldots, \mytt{X$_n^{\mathscr{X}_{j}}$}) are posted to them.
For each solver $\mathscr{X}_{j}$, new \mytt{X}$_i^{\mathscr{X}_{j}}$ variables are created, which form the equivalent $\mathscr{X}_{j}$ constraint.
%Then, $n \cdot m \cdot (m-1) / 2$ bridge constraints are created of the form \mytt{X$_i^{\mathscr{X}_{j}}$ \#==$_{\mathscr{X}_{j},\mathscr{X}_{k}}$ X$_i^{\mathscr{X}_{k}}$} ($1 \leq i \leq n$, $1 \leq j \leq m$, $j<m$), relating each pair of variables \mytt{X}$_i^{\mathscr{X}_{j}}$ and \mytt{X}$_i^{\mathscr{X}_{k}}$ once for each pair of different domains $\mathscr{X}_{j}$ and $\mathscr{X}_{k}$.
Then, $n \cdot m$ bridge constraints are created of the form \mytt{X$_i^{\mathscr{H}}$ \#==$_{\mathscr{H},\mathscr{X}_{j}}$ X$_i^{\mathscr{X}_{j}}$} ($1 \leq i \leq n$, $1 \leq j \leq m$), relating each pair of variables \mytt{X}$_i^{\mathscr{H}}$ and \mytt{X}$_i^{\mathscr{X}_{j}}$ once for each domain $\mathscr{X}_{j}$.
%
%
%  FOR THE FIGURE:
%
%ctr(\mytt{X$_1^{\mathscr{H}}$}, \ldots, \mytt{X$_n^{\mathscr{H}}$})
%
%ctr(\mytt{X$_1^{\mathscr{X}_1}$}, \ldots, \mytt{X$_n^{\mathscr{X}_1}$})
%
%ctr(\mytt{X$_1^{\mathscr{X}_m}$}, \ldots, \mytt{X$_n^{\mathscr{X}_m}$})
%
%\mytt{X$_1^{\mathscr{H}}$ \#==$_{\mathscr{H},\mathscr{X}_{1}}$ X$_1^{\mathscr{X}_{1}}$}
%
%\mytt{X$_n^{\mathscr{H}}$ \#==$_{\mathscr{H},\mathscr{X}_{1}}$ X$_n^{\mathscr{X}_{1}}$}
%
%\mytt{X$_1^{\mathscr{H}}$ \#==$_{\mathscr{H},\mathscr{X}_{m}}$ X$_1^{\mathscr{X}_{m}}$}
%
%\mytt{X$_n^{\mathscr{H}}$ \#==$_{\mathscr{H},\mathscr{X}_{m}}$ X$_n^{\mathscr{X}_{m}}$}

For example, let us consider a condition \mytt{X$_1$ > X$_2$} on the domain $\mathscr{H}$, and the compatible domains CLP($\mathscr{Q}$)
and CLP($\mathscr{FD}$).
The equivalent constraints in CLP($\mathscr{Q}$)
and CLP($\mathscr{FD}$) are respectively \mytt{clpq:\{X$_1^{\mathscr{Q}}$ > X$_2^{\mathscr{Q}}$\}}, and \mytt{X$_1^{\mathscr{FD}}$ \#> X$_2^{\mathscr{FD}}$} (following the concrete syntax of Prolog systems such as SICStus Prolog and SWI-Prolog, where \mytt{clpq:\{$C$\}} is the way to post the constraint $C$ to the CLP($\mathscr{Q}$) solver, and \mytt{\#>} is the $\mathscr{FD}$ operator corresponding to \mytt{>}).
Four bridges are built in this case: 
\mytt{X}$_1^\mathscr{H}$ \mytt{\#==}$_{\mathscr{H},\mathscr{Q}}$ \mytt{X}$_1^\mathscr{Q}$, 
\mytt{X}$_2^\mathscr{H}$ \mytt{\#==}$_{\mathscr{H},\mathscr{Q}}$ \mytt{X}$_2^\mathscr{Q}$,
\mytt{X}$_1^\mathscr{H}$ \mytt{\#==}$_{\mathscr{H},\mathscr{FD}}$ \mytt{X}$_1^\mathscr{FD}$, and
\mytt{X}$_2^\mathscr{H}$ \mytt{\#==}$_{\mathscr{H},\mathscr{FD}}$ \mytt{X}$_2^\mathscr{FD}$.

\subsection{Translation}
\label{sect:dl-to-clp-translation}

In this subsection, we show the translation of a Datalog program $\Pi_{DL}$ into a CLP program $\Pi_{CLP}$ such that $\Pi_{CLP}$ represents $\Pi_{DL}$.
We say that $\Pi_{CLP}$ represents $\Pi_{DL}$ if the meaning of $\Pi_{DL}$ is included in the meaning of $\Pi_{CLP}$ for any of its instance base relations.
For a Datalog program $\Pi_{DL}$, its meaning (denoted as $\llbracket \Pi_{DL} \rrbracket$) is the set of ground facts inferred for each relation.
For a CLP program, its meaning (denoted as $\llbracket \Pi_{CLP} \rrbracket$) is built from the set of all the (possibly non-ground) facts inferred for each relation: for each non-ground fact, all the type-compatible values constrained to the answer constraints are used to build the ground facts.
Thus, $\Pi_{CLP}$ represents $\Pi_{DL}$ if $\llbracket \Pi_{DL} \rrbracket \subseteq \llbracket \Pi_{CLP} \rrbracket$.

For example, the meaning of the Datalog program $\Pi_{DL} =$ \{\mytt{r(X):-X=1;X=2}\} is \{\mytt{r(1), r(2)}\} for a relation \mytt{r} that has integer type for its single argument.
The meaning of the CLP program (omitting domain annotations) $\Pi^1_{CLP} =$ \{\mytt{r(X):-X>0,X<3}\} is the same, provided the same integer type.
Note that, whereas non-recursive Datalog enjoys finite meanings (for finite relations), CLP can have infinite meanings as, e.g., $\Pi^2_{CLP} =$\{\mytt{r(X):-X>0}\}, whose meaning would be \{\mytt{r(1), r(2), r(3), ...}\}. 
%Note also that this program is not a valid Datalog program because of safety (which is a property that syntactically restricts Datalog programs to ensure finite meanings \cite{Ullman88}).
Both $\Pi^1_{CLP}$ and $\Pi^2_{CLP}$ represent the meaning of $\Pi_{DL}$, but the first one does it with a much better precision than the second one.
The program $\Pi^3_{CLP} =$ \{\mytt{r(X):-X>0,X<2}\} ($\llbracket \Pi^3_{CLP} \rrbracket =$ \{\mytt{r(1)}\}) does not represent $\Pi_{DL}$.

If a Datalog rule contains a call to a base relation (representing an SQL table), the translated CLP program omits that call to keep our approach instance-independent.
For example, the program \{\mytt{r(X):-t(X), X>17}\}, with a call to the base relation \mytt{t}, is translated into the CLP program \{\mytt{r(X):-true, X>17}\}.
Moreover, if the table has \mytt{CHECK} constraints, they are also added to the rule. 
Assuming that the declaration for this table is \mytt{CREATE TABLE t(a INT CHECK a>=0 AND a<=100)}, then the translation of the rule becomes: \mytt{r(X):-true, X>=0, X=<100, X>17}.

Next definitions formalize this translation from Datalog rules into CLP rules:

\begin{definition} \label{def:DLtoCLP}
The function \mytt{{\em DL\_to\_CLP}} takes a Datalog rule as input and returns a CLP rule.

\medskip

\mytt{{\em DL\_to\_CLP}}((\mytt{head :- goal$_1$, \ldots, goal$_n$})) = (\mytt{head :- goal$_1'$ , \ldots, goal$_n'$}) \\ \indent 
where \mytt{{\em DLGOAL\_to\_CLP}}(\mytt{goal$_i$}) = \mytt{goal$_i'$}\hfill $\square$

\end{definition}

\begin{definition} \label{def:DLGOALtoCLP}
The function \mytt{{\em DLGOAL\_to\_CLP}} takes a Datalog goal as input and returns a CLP goal.

\medskip

\mytt{{\em DLGOAL\_to\_CLP}}(\mytt{rel}) = \mytt{ctrs} \\ 
\indent where \mytt{rel} is a base relation, and \mytt{ctrs} is the conjunction of user-defined constraints for \mytt{rel}

\medskip

\mytt{{\em DLGOAL\_to\_CLP}}(\mytt{(goal$_1$ , goal$_2$)}) = \mytt{(goal$_1'$ , goal$_2'$)} \\ 
\indent where \mytt{{\em DLGOAL\_to\_CLP}}(\mytt{goal$_i$}) = \mytt{goal$_i'$}

\medskip

\mytt{{\em DLGOAL\_to\_CLP}}($meta$) = \mytt{goal$'$} \\ 
\indent where $meta$ is either \mytt{top($n$,goal)} or \mytt{distinct(goal)}, 
% or  \mytt{not goal}, \\ % not/1 is not handled up to now
%\indent 
and \mytt{{\em DLGOAL\_to\_CLP}}(\mytt{goal}) = \mytt{goal$'$}

\medskip

\mytt{{\em DLGOAL\_to\_CLP}}(\mytt{not(goal)}) = \mytt{true} 

\medskip

\mytt{{\em DLGOAL\_to\_CLP}}($exp_1 ~op~ exp_2$) = \mytt{ctr($exp_1 ~op~ exp_2$, $type$)} \\
\indent where $op$ is a comparison operator, and $type$ is the type of the expression

\medskip

The constraints and types for a given user-defined relation are taken from its metadata.
Relation types are used to annotate each logic variable in the program with its corresponding type.\hfill $\square$
\end{definition}

Note that a negated goal resulting from the translation of a relation defined by a statement such as \mytt{EXCEPT} restricts the meaning of the relation.
Here, we leave out this restriction because we do not deal with table instances.
Thus, the goal translation becomes simply \mytt{true}. 

\begin{proposition}
	\label{prop:translation}
The translation of a Datalog program $\Pi_{DL}$ into a CLP program $\Pi_{CLP}$ is a correct abstraction, i.e., $\llbracket \Pi_{DL} \rrbracket \subseteq \llbracket  \mytt{{\em DL\_to\_CLP}}(\Pi_{DL}) \rrbracket$.\hfill $\square$
\end{proposition}
The proof of this proposition is straightforward by checking that no case of Definition \ref{def:DLGOALtoCLP} removes solutions.

We assume a compatible mapping between SQL types and Datalog types.
From here on, we consider a Datalog type system consisting of the data types \mytt{string}, \mytt{integer} and \mytt{float}.

\begin{example}
    
%{\myfontcodesize
%    \begin{verbatim}
%    CREATE TABLE gas_products(
%    name      VARCHAR(20) PRIMARY KEY, 
%    butane    FLOAT CHECK butane    BETWEEN 0 AND 100, 
%    propane   FLOAT CHECK propane   BETWEEN 0 AND 100, 
%    olefins   FLOAT CHECK olefins   BETWEEN 0 AND 100, 
%    diolefins FLOAT CHECK diolefins BETWEEN 0 AND 100, 
%    CHECK     butane+propane+olefins+diolefins =  100);
%    
%    SELECT name FROM gas_products WHERE butane>60 AND propane>50;
%    \end{verbatim}
%}

Let us consider again the example of gas products presented in %Subsection \ref{sect:motivation}.
the introduction.
The result of the translation of the first SQL query into Datalog, followed by a simplification is:

{\myfontcodesize
\begin{verbatim}
{ (answer(N) :- gas_products(N,B,P,O,D), B>60, P>50) }
\end{verbatim}
}

Applying \mytt{{\em DL\_to\_CLP}} to this singleton, we get:

{\myfontcodesize
\begin{verbatim}
answer(N):-ctr(B>=0,float), ctr(B=<100,float), ctr(P>=0,float), ctr(P=<100,float),  
           ctr(O>=0,float), ctr(O=<100,float), ctr(D>=0,float), ctr(D=<100,float),  
           ctr(B+P+O+D=100,float), ctr(B>60,float), ctr(P>50,float).
\end{verbatim}
}

The first 9 constraints correspond to the \mytt{CHECK} constraints in the \mytt{CREATE TABLE} statement, whereas the last 2 constraints correspond to the conjunctive condition in the SQL query.
All the constraints have been annotated with the corresponding types declared for the table.\hfill $\square$

\end{example}

\subsection{Reasoning about Conditions}
\label{sect-clp-solving}

There are two cases to be considered: First, a CLP program resulting from the translation of an SQL query defining an $n$-ary relation $r$. 
In this case, the goal to test whether the query is consistent, inconsistent or simplifiable is $r(X_1,\ldots,X_n)$.
Second, a CLP program resulting from the translation of an SQL condition $c$, as those occurring in \mytt{CHECK} constraints for a given relation $r'$.
In this case, we build an SQL query of the form \mytt{SELECT * FROM $r'$ WHERE $c$} defining a fresh relation $r$, and we refer back to the first case.
In addition, we assume a function $solve(\phi, \Pi_{CLP})$ that takes a goal $\phi$ to be solved in the context of a logic program $\Pi_{CLP}$ and returns either a success substitution or failure.
This function represents the abstraction of the original Datalog program (next section briefly describes its implementation).
Only deterministic goals are considered in the analysis.

\begin{proposition}
	Given $\mytt{{\em DL\_to\_CLP}}(\Pi_{DL})=\Pi_{CLP}$, if $solve(\phi, \Pi_{CLP})=\bot$, then $solve(\phi, \Pi_{DL})=\bot$.\hfill $\square$
\end{proposition}
This proposition follows Proposition \ref{prop:translation} and states that if solving a goal $\phi$ for the logic program $\Pi_{CLP}$ that is an abstraction of another program $\Pi_{DL}$ leads to failure, then solving it for $\Pi_{DL}$ also leads to failure.
Thus, any inconsistent condition for an SQL query can be simply found by testing whether $solve(\phi, \Pi_{CLP})$ fails.
The first example in the introduction is an example of this:

$\phi$ = \mytt{answer(N,D,S)} \\ \indent
$\Pi_{CLP}$ = \{ (\mytt{answer(N,D,S) :- ctr(D='IT',string), ctr(D='HR',string)} \}

\noindent which fails, therefore identifying an inconsistent condition.

As well, a tautological condition can be found by complementing it and testing if $solve(\phi, \Pi_{CLP})$ fails.
An example of this is the table creation of employees (also in the introduction). 
For it, the following query is built and translated into a fresh relation \mytt{r} $\leftarrow$ \mytt{SELECT * FROM employees WHERE salary <= 2000 AND salary >= 5000}:

$\phi$ = \mytt{r(N,D,S)} \\ \indent
$\Pi_{CLP}$ = \{ (\mytt{r(N,D,S) :- ctr(S=<2000,integer), ctr(S>=5000,string)} \}

\noindent which also fails, therefore identifying a tautological condition.
Obviously, this procedure is not applied to true conditions, as it is the case of \mytt{WHERE}-less statements.

\begin{proposition}
Given $\mytt{{\em DL\_to\_CLP}}(\Pi_{DL})=\Pi_{CLP}$, if $solve(\phi, \Pi_{CLP})=\sigma$, and $solve(\phi, \Pi_{DL})=\theta$, then there exists $\eta$ such that $\theta=\eta\circ\sigma$.\hfill $\square$
\end{proposition}
This also follows Proposition \ref{prop:translation} and states that the success substitution $\sigma$ of solving $\phi$ for the logic program $\Pi_{CLP}$ is more general than $\theta$ (the one for $\Pi_{DL}$).
Thus, a simplifiable condition can be found by checking if any of the constrained variables in the program is bound after a successful CLP evaluation.
The last query in the introduction %Subsection \ref{sect:motivation} 
is an example.
Its translation is:

{%\setstretch{1}
$\phi$ = \mytt{answer(B,P)} \vspace*{-3mm}
\begin{tabbing}
\indent $\Pi_{CLP}$ = \{
(\=\mytt{answer(B,P) :-} \\[-1mm]
	\> \mytt{ctr(B>=0,float), ctr(B=<100,float),}  
	 \mytt{ctr(P>=0,float), ctr(P=<100,float),}  \\ [-1mm]
	\> \mytt{ctr(O>=0,float), ctr(O=<100,float),}   \mytt{ctr(D>=0,float), ctr(D=<100,float),}  \\[-1mm]
	\> \mytt{ctr(B+P+O+D=100,float),}
	 \mytt{ctr(B-P=10,float), ctr(B+P=80,float)}) \} 
\end{tabbing}
}

By solving this, the program is instantiated to:

{\setstretch{1}
\begin{tabbing}
$\Pi_{CLP}$ = \=\{
(\mytt{answer(45,35) :-} \\[-1mm]
    \> \mytt{ctr(45>=0,float), ctr(45=<100,float),}  
     \mytt{ctr(35>=0,float), ctr(35=<100,float),}  \\[-1mm]
    \> \mytt{ctr(O>=0,float), ctr(O=<100,float),}  
     \mytt{ctr(D>=0,float), ctr(D=<100,float),}  \\[-1mm]
    \> \mytt{ctr(45+35+O+D=100,float),}  
     \mytt{ctr(45-35=10,float), ctr(45+35=80,float)}) \}
\end{tabbing}
}

In this program, ground conditions can be replaced by true conditions (because it has been proven by $solve$ that they succeed). 
Therefore, they are simplifiable and a warning can be raised.

\section{System Implementation}
\label{sect:the-system}

In this section, a system implementing the proposed approach to SQL semantic error identification is described.
We developed this proposal in the deductive database system DES (Datalog Educational System, \url{des.sourceforge.net}) version 6.0.

\subsection{Defining Solver Cooperation}

In the current implementation, there is opportunity for the cooperation of the exact numerical solvers  $\mathscr{Q}$ and $\mathscr{FD}$.
To implement this, we define the way for solving an integer constraint, which follows the approach described in Section \ref{sect:cooperation}.
The following code excerpt shows posting an integer constraint to both solvers  $\mathscr{Q}$ and $\mathscr{FD}$ (the case of a single solver is a simplification of this predicate and its description is thus omitted):

{\myfontcodesize
\begin{verbatim}
1. post_clp_ctr(ctr(Cond,integer),InputBridges,OutputBridges) :-
2.   Cond =.. [Op,L,R], copy_term([L,R],[FDL,FDR]), copy_term([L,R],[QL,QR]),
3.   term_variables([L,R],Xs),
4.   term_variables([LFD,RFD],XFDs), term_variables([LQ,RQ],XQs),
5.   op_fdop(Op,FDOp),
6.   CtrFD =.. [FDOp,LFD,RFD], CtrQ  =.. [Op,LQ,RQ],
7.   add_bridges(fd,Xs,XFDs,InputBridges,Bridges),
8.   add_bridges(q,Xs,XQs,Bridges,OutputBridges),
9.   catch(call(CtrFD),_,true), catch(clpq:{CtrQ},_,true).
\end{verbatim}
}

This predicate has two input arguments and a third output argument.
The first one is the constraint, the second one is the list of already built bridges along solving, and the third one is for the output bridges (possibly augmenting the input bridges with new ones).
Line 2 identifies the condition with its left and right arguments (\mytt{L} and \mytt{R}, respectively) and makes a copy (to be sent later on) as part of the $\mathscr{Q}$ and $\mathscr{FD}$ constraints, which are built in line 6, and posted to the corresponding solvers in line 9.
The CLP operator for the domain $\mathscr{Q}$ is the same as for Datalog, but its correspondence with the $\mathscr{FD}$ operator has to be found (line 5) to build a syntactically correct $\mathscr{FD}$ constraint.
Lines 7--8 build the bridges between each variable \mytt{X}$^\mathscr{H}$ in the original condition and its counterpart variables \mytt{X}$^\mathscr{Q}$ and \mytt{X}$^\mathscr{FD}$.
Because of variable sharing, it may be the case that a bridge for a given variable has been previously built.
All the previous bridges are stored in the input list \mytt{InputBridges} and no bridge is added by the predicate \mytt{add\_bridges}/5 if already present in this list.
This predicate has as input arguments the domain for which to build the bridge, the variables in the condition, the copy of these variables in the constraint domain, and the input bridges.
Its last argument (\mytt{OutputBridges}) will contain the input bridges plus the new one (if eventually created).
For each variable in \mytt{Xs}, it calls \mytt{add\_bridge}/3, which adds a new bridge if needed:

{\myfontcodesize
\begin{verbatim}
% Existing bridge: just retrieve bindings
add_bridge(bridge(D,X,Y),Bridges,Bridges) :- bridge_in(bridge(D,X,Y),Bridges), !.
% New bridge: add it to the output list
add_bridge(bridge(D,X,Y),Bridges,[bridge(D,X,Y)|Bridges]) :-
  add_domain_binding_daemon(D,X,Y).
\end{verbatim}
}

Here, a bridge has the form \mytt{bridge(D,X,Y)}, where \mytt{D} is the domain, \mytt{X} is the variable in the domain $\mathscr{H}$, and \mytt{Y} is the variable in the domain \mytt{D} (this term corresponds to the previous notation \mytt{X$^{\mathscr{H}}$ \#==$_{\mathscr{H},\mathscr{D}}$ Y$^{\mathscr{D}}$}).
The predicate \mytt{bridge\_in}/2 checks if the input bridge is already built and, if so, it retrieves the bindings for both variables.
Otherwise, a new bridge is created:

{\myfontcodesize
\begin{verbatim}
bridge_in(bridge(D,X,Y),[bridge(D,BX,Y)|_Bri]) :- var(BX), X==BX, !.
bridge_in(bridge(D,X,Y),[_|Bri]) :- bridge_in(bridge(D,X,Y),Bri).
\end{verbatim}
}

Finally, the predicate \mytt{add\_domain\_binding\_daemon} creates a daemon %(with the help of the Prolog built-in \mytt{freeze}, 
(which suspends the goal in its second argument until its first argument becomes ground).
It is activated by grounding either \mytt{X} in $\mathscr{H}$ or \mytt{Y} in \mytt{D}: % (recall that each domain requires a different syntax for posting a constraint):

{\myfontcodesize
	\begin{verbatim}
add_domain_binding_daemon(fd,X,FD):- freeze(X,FD#=X), freeze(FD,X=FD).
add_domain_binding_daemon(q,X,Q):- freeze(X,clpq:{Q=X}), freeze(Q,q_to_int(Q,X)).
\end{verbatim}
}

The call to the predicate \mytt{q\_to\_int} converts compatible numbers between rationals and integers.

The function $solve(\phi, \Pi_{CLP})$ has been implemented with a CLP metainterpreter written in Prolog.
The predicate for to this function is \mytt{clp\_evaluation(Goal, Program, InputBridges, OutputBridges)}, where \mytt{Goal} is the argument corresponding to $\phi$, and \mytt{Program} corresponds to $\Pi_{CLP}$.
The arguments \mytt{InputBridges} and \mytt{OutputBridges} are added to keep track of the bridges being created during solving.
The sketch of this predicate is similar to a Prolog metainterpreter \cite{Sterling:1994:APA:175753}, but when a constraint term is identified as a goal, it calls the predicate \mytt{post\_clp\_ctr}/3.

\subsection{A System Session}

A system session log for the examples in the introduction can be found at {\myfontcodesize\url{www.fdi.ucm.es/profesor/fernan/DES/prole2018/examples.pdf}}, which we omit here for the sake of space.
In addition to these examples, note that our approach is applied to conditions as complex as needed, including subqueries.
For example, subqueries in expressions are allowed:

{\myfontcodesize
	\begin{verbatim}
DES> SELECT (SELECT ename FROM employees WHERE salary BETWEEN 5000 AND 1000) 
     FROM departaments WHERE dname='Human resources';
Warning: Inconsistent condition.
Warning: Missing join condition for [departments,employees].
\end{verbatim}
}

The next condition includes a subquery in the \mytt{WHERE} clause and is inconsistent:

{\myfontcodesize
	\begin{verbatim}
DES> SELECT ename FROM employees
     WHERE salary<1000 AND salary>(SELECT salary FROM employees WHERE salary>2000);
Warning: Inconsistent condition.
\end{verbatim}
}

Note that inconsistency in this case is due to the combination of the conditions of both the root query and its subquery.
The translations are:

$\Pi_{DL}$=\{(\mytt{answer(A) :- employees(A,\_B,C), employees(\_D,\_E,F), F>2000, C>F, C<1000})\}

$\Pi_{CLP}$=\{(\mytt{answer(A) :- ctr(F>2000,integer), ctr(C>F,integer), ctr(C<1000,integer)})\}

\noindent where it can be seen that the last three constraints cannot be fulfilled.

\subsection{Performance}

%An interesting point to be clarified is whether the proposed method is enough performant to deal 
This section describes the performance of our approach when dealing
with queries of relevant size.
Our experience at classroom indicates that the tool successfully handles students' queries (including long queries for solving complex SQL puzzles which were posed to outstanding students \cite{sae19b-prole19}).
However, the tool might be used to verify very long queries which are automatically generated by other tools (e.g., handling of persistence in Object-Relational Mapping approaches such as Hibernate ORM).
Thus, this section analyses the cost of an SQL query translation (i.e., the function \mytt{{\em SQL\_to\_DL}} to translate SQL into Datalog in Subsection \ref{sect:sql-to-dl-translation}, and the function \mytt{{\em DL\_to\_CLP}} to translate Datalog into CLP in Subsection \ref{sect:dl-to-clp-translation}), and CLP solving (the function $solve$ in Subsection \ref{sect-clp-solving}).
For the experiments we have selected a database with $n$ empty tables, each one with a column with a \mytt{CHECK} constraint over integers (so that solver cooperation is applied), and a query $Q_1$ consisting of $n-1$ nested subqueries, inductively defined as:

\begin{tabbing}
	$Q_n$\=$\equiv$ \=\mytt{SELECT t$_n$.a} \=\mytt{FROM t$_n$} \=\mytt{WHERE t$_n$.a>$n$}\\
	%\indent
	$Q_i$\>$\equiv$ \>\mytt{SELECT t$_i$.a} \>\mytt{FROM t$_i$} \>\mytt{WHERE t$_i$.a>$i$ AND t$i$.a IN ($Q_{i+1}$)}
\end{tabbing}

\setlength{\tabcolsep}{3pt}
\begin{wrapfigure}{r}{8cm}
\vspace{-0.7cm}
	\centering
	{\small
	\begin{tabular*}{8cm}{rrrrrrr}
		\hline
		$n$ & \mytt{{\em SQL\_to\_DL}} & \mytt{{\em DL\_to\_CLP}} & $solve$ & Total & DB2 & Oracle \\
		\hline 
		 10 &     3 &	0 &	1 &	   65 &    169 &    111 \\
		 20 &    11 &	0 &	1 &	  118 &    340 &    200 \\
		 30 &    28 &	0 &	1 &	  196 &    607 &    631 \\
		 40 &    63 &	1 &	2 &	  314 &  2,135 &  1,821 \\
		 50 &   121 &	1 &	2 &	  510 &  4,278 &  4,425 \\
		 60 &   205 &	1 &	3 &	  753 &  7,990 &  9,292 \\
		 70 &   307 &	2 &	4 &	1,113 & 15,298 & 18,049 \\
		 80 &   438 &	1 &	4 &	1,491 & 28,288 & 30,215 \\
		 90 &   633 &	2 &	4 &	2,050 & 48,314 & 50,350 \\
		100 & 1,077 &	2 &	7 &	2,639 & 96,139 & 78,292 \\
		\hline
	\end{tabular*}
}
\vspace{-0.5cm}
	%\rule{3cm}{7cm}
\end{wrapfigure}
For a given $n$, $Q_1$ is the test query involving $n$ correlated base tables, and inheriting the constraints in each table definition.
The table to the right shows times in milliseconds (got with \mytt{statistics/2} for walltime) for the analysed steps.
The column `Total' lists the total run time (this includes other tasks such as parsing and processing).
In addition, for the sake of comparing this total with those of well-known relational database systems, the two final columns `DB2' and `Oracle' show the total run time for IBM DB2 version 11.1.0 and Oracle version 11g via an ODBC connection from the tool.
As a test platform, we used a Windows 10 64-bits OS running on an Intel Xeon CPU E3-1505M v5 (4 physical cores) running at 2.8 GHz, with 16GiB RAM. 
We used the source distribution of DES version 6.2 (with a bit of extra code for measuring time) running on SICStus Prolog 4.4.1 64-bits.
As expected, both translating Datalog to CLP, and solving the CLP program take negligible time, whereas the translation from SQL to Datalog takes a reasonable time.
Note that \mytt{{\em SQL\_to\_DL}} includes in particular program transformations (folding/unfolding), safety checks, argument mode handling, and simplifications.
Both DB2 and Oracle do not seem to scale well with this kind of queries.
Other database systems at hand (MySQL 5.7.13 and MS SQL Server 2014) could not handle so many nested levels (with a limit of 64 and 18, respectively).
%Figure \ref{table-performance}
%

\subsection{Supported Semantic Errors}
\label{sect:errors}

This section lists and briefly describes our approach to deal with all the supported semantic errors (identified by numbers in \cite{Brass:2006:SES:1183058.1183064}) in our implementation.
The analysis incorporates the bindings produced along a successful CLP program solving.

\begin{itemize}
\item Error 1: Inconsistent condition. 
If the evaluation of the CLP program fails, a warning is issued.
This can be easily extended to display the source condition corresponding to the failing constraint by annotating each constraint with its corresponding condition.
\item Error 2: Unnecessary \mytt{DISTINCT}. 
A warning is issued if the query returns no duplicates and includes this modifier with respect to the primary keys in the involved relations.
\item Error 3: Constant output column.
As a consequence of CLP solving, a column can become ground.
\item Error 4: Duplicated column values.
Two or more columns can be assigned to the same logical variable representing its output.
\item Error 5: Unused tuple variable.
An unaccessed single relation in the FROM list from the root query (Error 27 captures all other cases).
\item Error 6: Unnecessary join.
Check if no column in a join is used in addition to its correlation, if any.
Foreign keys are taken into account, otherwise, false positives might be raised.
\item Error 7: Tuple variables are always identical.
A warning is issued if two or more relations produce the same tuples.
This is accomplished by testing if the same goal occurs more than once with the same variables.
\item Error 8: Implied or tautological condition. 
The original Error 8 included an inconsistent condition, which is checked in Error 1 above.
Checking this is based on testing whether the complement of the condition fails,
meaning that the condition is trivially true.
\item Error 9: Comparison with \mytt{NULL}. 
This is performed in the SQL syntax tree by looking for comparisons with null values.
\item Error 11: Unnecessary general comparison operator.
A warning is issued if \mytt{LIKE '\%'} occurs, which is equivalent to \mytt{IS NOT NULL} by inspecting the SQL syntax tree.
Additionally it issues a warning about trivially true (resp. false) conditions as \mytt{cte LIKE '\%'} (resp. \mytt{NOT LIKE}).
This might also be checked by a string solver in Error 1.
\item Error 12: \mytt{LIKE} without wildcards.
Again, this error is straightforwardly checked by inspecting the SQL syntax tree.
\item Error 13: Unnecessarily complicated \mytt{SELECT} in \mytt{EXISTS}-subquery.
Detect patterns different from \mytt{SELECT *} as the root in an existential subquery.
\item Error 16: Unnecessary DISTINCT in aggregation function.
A warning is issued if either \mytt{MIN} or \mytt{MAX} is used with a \mytt{DISTINCT} argument, as well as if other aggregate is used with a \mytt{DISTINCT} expression involving key columns.
In both cases, the Datalog translation is inspected.
\item Error 17: Unnecessary argument of \mytt{COUNT}.
A warning is issued if \mytt{COUNT} is applied to an argument that cannot be null as a primary key.
Metadata is used to determine non-null arguments.
\item Error 27: Missing join condition.
A warning is issued if two relations are not joined by a criterium.
This includes Error 5 for a single unused relation.
\item Error 32: Strange \mytt{HAVING}.
A warning is issued if a \mytt{SELECT} with \mytt{HAVING} does not include a \mytt{GROUP BY} by inspecting the SQL syntax tree.
\item Error 33: \mytt{SUM(DISTINCT ...)} or \mytt{AVG(DISTINCT ...)}. 
A warning is issued if duplicate elimination is included for the argument of either \mytt{SUM} or \mytt{AVG}.
If included, this might not be an error, but it is suspicious because duplicates are usually relevant for these aggregates.

\end{itemize}

\section{Related Work}
\label{sect:related-work}

There are many tools targeted at learning SQL focusing on the answers of queries with respect to database instances, i.e., comparing the output of queries with the output of reference queries provided by experts 
(e.g.,
%ACME \cite{soler2006web}, 
SQLator \cite{Sadiq:2004:SOS:1026487.1008055}, 
WebSQL \cite{Allen2000WebSQLAI}, 
AsseSQL \cite{Prior:2014:AOB:2591708.2602682}, 
%SQLZoo\footnote{\url{https://sqlzoo.net}},
%HackerRank\footnote{\url{https://www.hackerrank.com/}},
%LearnSQL % No está disponible en abierto, no se encuentran referencias.
%SQLJudge \cite{SQLJudge}, 
and
QueryViz \cite{DBLP:journals/pvldb/Gatterbauer11}).
Other set of tools are focused on the semantic aspects of queries as SQL Tutor \cite{Mitrovic2012} which uses Constraint-Based Modelling (CBM) to form models of its students, enabling the automatic selection of problems based on these models.
Another one is SQL-LTM \cite{Dollinger20103i}, a tutoring module relying on reference queries to which student queries are compared.

The semantic-based system \mytt{sqllint} \cite{Brass:2006:SES:1183058.1183064} is the closest approach to ours. 
They introduce the concept of soft keys (attributes used in practice for identifying tuples, but that can have duplicates; e.g., the name of a person) which would be useful for some of their checks.
A description of the way to identify such errors is given in \cite{1579136}, relying on \textit{ad-hoc} consistency checks somewhat based on classical techniques \cite{Guo1996SolvingSA}.
With respect to subqueries, it only supports \mytt{EXISTS}  (no \mytt{IN}, \mytt{>= ALL}, \ldots)
Their approach neither supports aggregates, nor \mytt{UNION}, nor \mytt{LIKE}, and nor \mytt{IS [NOT] NULL}.
As types, it includes only strings and integers, % constants, and no other numeric constants.
and expressions are not allowed.
Finally, it does not support \mytt{CHECK} constraints in table definitions.
%Both \mytt{sqllint} and DES systems deal with error numbers 1, 2, 3, 4, 5, 8 and 27; in addition to this, \mytt{sqllint} deals with errors 34 and 39, while DES deals with errors 6, 7, 9, 11, 12, 13, 16, 17, 32, and 33.
Note also that, in contrast to DES, \mytt{sqllint} is only an analyzer, not a complete SQL system with a solving engine which could be used for teaching.

\section{Conclusions and Future Work}
\label{sect:conclusions}

We have presented a system using constraint logic programming for the semantic analysis of SQL statements (both DML and DDL).
With the aim of detecting possible misuses of syntactically correct SQL statements at compile-time, this system focuses on both metadata and statements, instead of data from tables.
There have been other approaches to SQL analysis (targeted at comparing results for concrete database instances, and based on CBM techniques), but ours mainly follows the same path as \mytt{sqllint}.
However, instead of using consistency techniques as in that work, we use CLP constraints and solver cooperation to develop a precise analysis, which can deal with non-linear conditions and queries as complex as needed.
Reasoning at the logic level eases the development of this approach, instead of using the more cumbersome SQL formulations for consistency checking.
Performance data show that the approach is practical, and well able to cope with queries that other systems cannot afford.

Despite we have successfully evaluated the tool in classroom and students have appreciated the semantic feedback, a more thorough evaluation must be done.
As part of a teaching innovation project, we are currently analysing the tool with both on-line questionnaires provided to students, and logging user sessions.
While questionnaires include selectable answers in a Likert scale (and also open answers to express additional specific comments), logs can be inspected to observe the reaction of students to the semantic warnings.
In addition, there is ample room for future work as, for example, the development (most likely, with CHR) of specific solvers for types such as strings and dates (in addition to the already used, but simpler, domain $\mathscr{H}$).
Taking into account bindings and domain pruning in negated CLP goals resulting from the translation of constructs such as \mytt{NOT IN} and \mytt{NOT EXISTS} would also increase the precision of the analysis.
Also, the tool might propose simplified versions of conditions by decompiling the CLP constraint store into SQL.
Other potential additions include: a Boolean type and its handling with a CLP($\mathscr{B}$) solver; taking advantage of subtypes (exact numeric types with limited range, and string subtypes with bounded size); and implementing projections \cite{DBLP:journals/corr/abs-0904-2136}.
Finally, the proposal in this work can be applied to the semantic analysis of Datalog queries and programs, and to other scenarios such as the verification of automatically-generated SQL queries.
%Finally, we plan to evaluate the tool in classroom by automatic analysis of student system session logs.

\bibliographystyle{acmtrans}
%\bibliography{biblio}

\end{document}